\documentclass[12 pt, amsfonts,amsmath, amssymb,color]{article}

\usepackage{amsmath,amssymb,amsfonts,latexsym,float,graphics,epsfig}
\usepackage{subfig}
\usepackage{verbatim}
\usepackage{relsize}
\usepackage{hyperref}

\def\be{\begin{equation}}
\def\ee{\end{equation}}
\def\bea{\begin{eqnarray}}
\def\eea{\end{eqnarray}}

\begin{document}
%<<<<<<<<<<< enumeration of eqns section wise>>>>>>>>>>>>>>>>>>>

\renewcommand\theequation{\arabic{section}.\arabic{equation}}
\catcode`@=11 \@addtoreset{equation}{section}
%<<<<<<<<<<<<<<<<<<<<<<<<<<<<<<<<<>>>>>>>>>>>>>>>>>>>>>>>>>>>>>>>>>
\newtheorem{axiom}{Definition}[section]
\newtheorem{theorem}{Theorem}[section]
\newtheorem{axiom2}{Example}[section]
\newtheorem{lem}{Lemma}[section]
\newtheorem{prop}{Proposition}[section]
\newtheorem{cor}{Corollary}[section]

\newcommand{\ben}{\begin{equation*}}
\newcommand{\een}{\end{equation*}}
\title{\bf Thermodynamic geometry and interacting microstructures of BTZ black holes}
\author{
\bf Aritra Ghosh\footnote{E-mail: ag34@iitbbs.ac.in} \hspace{0.5mm} and Chandrasekhar Bhamidipati\footnote{E-mail: chandrasekhar@iitbbs.ac.in} \\
~~~~~\\
 School of Basic Sciences, Indian Institute of Technology Bhubaneswar,\\   Jatni, Khurda, Odisha, 752050, India\\
}

\date{ }

\maketitle

\begin{abstract}
In this work, we present a study to probe the nature of interactions between black hole microstructures for the case of the BTZ black holes. Even though BTZ black holes without any angular momentum or electric charge thermodynamically behave as an ideal gas, i.e. with non-interacting microstructures; in the presence of electric charge or angular momentum, BTZ black holes are associated with repulsive interactions among the microstructures. We extend the study to the case of exotic BTZ black holes with mass $M = \alpha m + \gamma \frac{j}{l}$ and angular momentum $J=\alpha j + \gamma l m$, for arbitrary values of $ (\alpha, \gamma)$ ranging from purely exotic $(\alpha=0,\gamma=1)$, slightly exotic $(\alpha > \frac{1}{2},\gamma < \frac{1}{2})$ and highly exotic $(\alpha < \frac{1}{2}, \gamma > \frac{1}{2})$. We find that unlike the normal BTZ black holes (the case $(\alpha =1,\gamma =0)$), there exist both attraction as well as repulsion dominated regions in all the cases of exotic BTZ black holes.

\end{abstract}

\smallskip

% \maketitle
\section{Introduction}
Since the work of Bekenstein \cite{Bekenstein:1973ur,Bekenstein:1974ax}, Hawking \cite{Hawking:1974sw,Hawking:1976de} and related developments \cite{Bardeen:1973gs,Gibbons:1976ue}, there have been active efforts to understand the microscopic degrees of freedom, thermodynamics and phase transitions \cite{Hawking:1982dh} of black holes, more recently, in extended thermodynamic phase space~ \cite{Henneaux:1984ji}-\cite{KB}.
Further, there has been a considerable amount of interest in the study of thermodynamic geometry of black holes and their underlying microstructures \cite{Wei:2015iwa} - \cite{GBRuppeiner}. Since it is possible to define a temperature for a black hole, it is then the most natural to think of an associated microscopic structure. These black hole microstructures are in general interacting just like the molecules in a non-ideal fluid. The fact that asymptotically AdS black holes have thermodynamic behavior similar to that of a van der Waals fluid \cite{Chamblin:1999tk,Chamblin:1999hg,Kubiznak:2012wp} makes the picture of fluid like interacting microstructures quite natural. The commonly adapted technique to probe the nature of interactions between these microstructures is to look at the thermodynamic geometry of the corresponding macroscopic system. For example, the study of Ruppeiner geometry in standard thermodynamics reveals that for a particular system, the curvature of the Ruppeiner metric indicates the nature of interactions between the underlying molecules. For a system where the microstructures interact attractively as in a van der Waals fluid, the curvature scalar of the Ruppeiner metric carries a negative sign whereas in a system where the microstructures interact in a repulsive manner, the curvature is positive. For a non-interacting system such as the ideal gas, the metric is flat. In the attempts to probe the nature of interactions between the microstructures, Ruppeiner geometry has therefore been applied quite extensively in black hole thermodynamics. The investigation of the nature of interactions between the black hole microstructures was first done in the context of Banados, Teitelboim and Zanelli (BTZ) black holes in three dimensions in~\cite{Cai:1998ep}. Furthermore in~\cite{Shen:2005nu}, Ruppeiner geometry for Reissner-Nordstr\"{o}m, Kerr and Reissner-Nordstr\"{o}m-AdS black holes was explored in the non-extended thermodynamic space, where the divergence of the scalar curvature is consistent with the Davies' phase transition point~\cite{Davies:1978mf}.
For example, for the case of both Reissner-Nordstr\"{o}m-AdS (RN-AdS) and electrically charged Gauss-Bonnet-AdS (GB-AdS) black holes \cite{Wei:2015iwa,AR,Mann2019,Wei:2019yvs,GBRuppeiner}, there is a competition between attractive and repulsive interactions between the microstructures of the black hole. In the case where attraction exactly balances the repulsion, the black hole can be regarded as effectively non-interacting. It has also been recently noted that the Schwarzschild-AdS \cite{SAdS} and neutral GB-AdS black holes \cite{Wei:2019ctz} are dominated by attractive interactions among the putative microstructures.\\

\noindent

\textbf{Motivation and results:} In this work, with a motivation to probe the nature of interactions among the black hole microstructures, we study BTZ black holes in the extended phase space with full generality including the effects due to electric charge and angular momentum. In the non-extended thermodynamic phase space, Ruppeiner geometry has been applied earlier to study BTZ black holes in various approaches \cite{Cai:1998ep,Sarkar:2006tg,Quevedo:2008ry,Akbar:2011qw,Hendi:2015fya,Mohammadzadeh:2018zod}. However, to decipher the true microstructures of black holes in AdS and to know the regions of attraction and repulsive behavior, a thorough study in extended phase space (where the fluidlike behavior is most clear due to the availability of an equation of state) is required, as noted recently in~\cite{Wei:2015iwa,Mann2019}. In three dimensions, there are interesting possible connections with exact formulas available to study the statistical interpretations from the holographically dual side \cite{Brown:1986nw,Strominger:1997eq,Birmingham:1998jt}.
Motivated by this, we study these \((2+1)\)-dimensional black holes, having a thermodynamic behavior which is not only interesting in itself but also provides clues to the understanding of the higher dimensional counterparts. For example, the thermodynamics of the rotating BTZ black hole is much more transparent as compared to that of a Kerr black hole in higher dimensions. In other words, these \((2+1)\)-dimensional black holes, even though simpler to study can be regarded as toy models that can be used to shed light into the behavior of black holes in \(d \geq 4\). We show that a charged and non-rotating BTZ black hole is dominated by repulsive interactions. This is expected since it is now understood that charged microstructures interact in a repulsive manner \cite{Wei:2015iwa,AR,Mann2019,Wei:2019yvs,GBRuppeiner}. We also note that a neutral but rotating BTZ black hole behaves qualitatively in a similar manner as a charged BTZ and is dominated by repulsive interactions. This is because at the thermodynamic level, the angular momentum of the black hole behaves very much like electric charge and therefore from the phenomenological level, it suggestive to associate microstructures to rotating BTZ black holes that interact repulsively. For general case which includes both rotation and electric charge, these two repulsive effects can add up and therefore there is no attraction at all.

We also study the case of exotic BTZ black holes \cite{Townsend:2013ela,Frassino:2015oca,Frassino:2019fgr}, where the roles of mass and angular momentum are reversed \cite{Carlip:1991zk,Carlip:1994hq,Banados:1997df,Banados:1998dc} and in the general case, the solutions may emerge from a gravitational action \cite{Deser:1981wh,VanNieuwenhuizen:1985ff,Witten:1988hc,Horne:1988jf,Afshar:2011qw}, which is a linear combination of the standard Einstein-Hilbert, together with the Chern-Simons action \cite{Deser:1981wh}. The study of holography with gravitational Chern-Simons terms has opened up interesting new avenues \cite{Iyer:1994ys,Jacobson:1994qe,Kraus:2005vz,Kraus:2005zm,Maldacena:1998bw,Solodukhin:2005ah,Sahoo:2006vz,Sen:2007qy}.
For the exotic BTZ black hole case \cite{Townsend:2013ela,Frassino:2015oca,Frassino:2019fgr}, we find that the Ruppeiner curvature scalar changes sign with the inner horizon radius. Therefore, there are both attraction and repulsion dominated regions for exotic BTZ black holes, in general.

The paper is organized as follows. In the next section we give a very brief review of thermodynamic geometry that will be useful for the rest of the paper. We begin our study of Ruppeiner geometry for BTZ black holes in section 3 starting with the general charged and rotating case hence discussing the special cases. The results then physically interpreted from a non-ideal fluid point of view. In section 4, we explore the thermodynamic geometry of neutral exotic BTZ black holes and show that there are both attraction and repulsion dominated regions as compared to only repulsive interactions for the ordinary BTZ black holes with charge and/or angular momentum. We end the paper with remarks in section 5.

\section{Thermodynamic geometry}
In this section, we recall the essential aspects of the geometry of the thermodynamic phase space and how it naturally leads to Ruppeiner geometry. It is well known that the thermodynamic phase space is endowed with a contact structure \cite{hermann} - \cite{bravetti2015}, i.e. it assumes the structure of a contact manifold. This means that the thermodynamic phase space is a pair \((\mathcal{M},\eta)\) where \(\mathcal{M}\) is a manifold of odd dimension (say \((2n+1)\)) that is smooth and that \(\mathcal{M}\) is equipped with a one form \(\eta\) such that the volume form, \(\eta \wedge (d\eta)^n\) is non-zero everywhere on \(\mathcal{M}\). One can always locally find (Darboux) coordinates \((s,q^i,p_i)\) on \(\mathcal{M}\) where \(q^i\) and \(p_i\) are said to be a conjugate pairs such that,
\begin{equation}\label{eta}
  \eta = ds - p_idq^i.
\end{equation}
Of interest in thermodynamics are a very special class of submanifolds of \(\mathcal{M}\), known as the Legendre submanifolds. If \(\Phi:L \rightarrow \mathcal{M}\) be a submanifold with \(\Phi\) being an embedding, then if \(\Phi^*\eta = 0\) or equivalently, \(\eta_L = 0\), one calls \(L\) an isotropic submanifold. From eqn (\ref{eta}), it is clear that an isotropic submanifold cannot include a conjugate pair of local coordinates. Further, if \(L\) is an isotropic submanifold of maximal dimension, it is then a Legendre submanifold. It is easy to verify that the maximal dimension is \(n\) and therefore, all Legendre submanifolds are \(n\)-dimensional. On an arbitrary Legendre submanifold \(L\), one has,
\begin{equation}\label{etavanish}
  ds - p_idq^i=0.
\end{equation}
One immediately identifies this statement as a first law of thermodynamics which for black holes in the extended phase space is given by,
\begin{equation}\label{firstlaw}
  dM - TdS - VdP - \Phi dQ - \Omega dJ = 0,
\end{equation}
with the identifications \(s = M, q^1 = S, q^2 = P, q^3 = Q, q^4=\Omega, p_1 = T, p_2 = V, p_3 = \Phi, p_4=J\). Here, \(M\) is the mass of the black hole which in the extended thermodynamics framework is equated to the enthalpy, i.e. \(M = H\) while other symbols\footnote{In this paper, we shall work in the fixed charge \(Q\) and angular momentum \(J\) ensemble, i.e. we set \(dQ = dJ = 0\) in subsequent discussions.} have their usual meanings from black hole thermodynamics.

\smallskip

A contact manifold can be associated with a Riemannian metric (see \cite{therm} - \cite{mrugala1996}) \(G\) which is in a sense compatible with the contact structure. The metric is bilinear, symmetric and non-degenerate. It is usually written down in the coordinate form as,
\begin{equation}G = \eta^2 - dp_idq^i.\end{equation}
However, the full metric is not particularly as important in thermodynamics as is its projections on various Legendre submanifolds. Legendre submanifolds appearing in thermodynamics have the local form given by,
\begin{equation}\label{thermodynamiclocal}
  s = \phi(q^i); \hspace{3mm} p_i = \frac{\partial \phi(q^i)}{\partial q^i},
\end{equation}
where \(\phi = \phi(q^i)\) is a thermodynamic potential and is known as the generator of the Legendre submanifold (say \(L\)). It then follows that when projected on \(L\), the metric takes the form,
\begin{equation}\label{GLlocal}
  G|_L = -dp_idq^i|_L = -\frac{\partial^2 \phi}{\partial q^i \partial q^j}dq^idq^j; \hspace{5mm} i,j \in \{1,2,....,n\}.
\end{equation}
This exactly corresponds to the metric of Weinhold \cite{Weinhold} and Ruppeiner \cite{Ruppeiner} whose lines elements are respectively given as,
\begin{equation}\label{Weinhold-Ruppeiner}
  ds^2_W =  -\frac{\partial^2 U}{\partial x^i \partial x^j}dx^idx^j; \hspace{5mm} ds^2_R = -\frac{\partial^2 S}{\partial x^i \partial x^j}dx^idx^j,
\end{equation} where \(\{x^i\}\) are independent thermodynamic variables. It is a simple exercise to show that \(ds^2_R = \beta ds^2_W\) where \(\beta = 1/T\) is the inverse temperature so the metrics differ only by a conformal factor. In case of static black holes however, since entropy and volume are not independent thermodynamic variables, the study of metric structures should not be done taking the internal energy \(U =U(S,V)\) as a fundamental potential \cite{SAdS}.

\smallskip

In the extended thermodynamic phase space framework, taking the enthalpy (equated to the mass of the black hole) as the generator of the Legendre submanifold representing the black hole, it follows that on the \((S,P)\)-plane, the Ruppeiner line element after some manipulations can be written down in the form \cite{SAdS,GBRuppeiner},
\begin{equation}\label{RuppeinerSP}
  ds_R^2 = \frac{1}{C_P}dS^2 + \frac{2}{T}\bigg(\frac{\partial T}{\partial P}\bigg)_SdSdP - \frac{V}{TB_S}dP^2,
\end{equation}
where \(C_P\) is the specific heat at constant pressure and \(B_S = -V(\partial P/\partial V)_S\) is the adiabatic bulk modulus of the black hole. Notice that \(B_S = \infty \) for black holes where \(V\) and \(S\) are not independent and the last term drops from eqn (\ref{RuppeinerSP}). Alternatively, the Ruppeiner metric is also often calculated on the \((T,V)\)-plane in a representation where the fundamental thermodynamic potential is naturally the Helmholtz potential. The line element for the general case is given by \cite{SAdS,GBRuppeiner},
\begin{equation}\label{RuppeinerTV}
  ds_R^2 = \frac{1}{T}\bigg(\frac{\partial P}{\partial V}\bigg)_TdV^2 + \frac{2}{T}\bigg(\frac{\partial P}{\partial T}\bigg)_VdTdV + \frac{C_V}{T^2}dT^2,
\end{equation}
where \(C_V\) is the specific heat at constant volume, which turns out to be zero for black holes where the thermodynamic volume and the entropy are not independent. The scalar curvature remains equivalent on both the \((S,P)\)- and \((T,V)\)-planes. This can be easily verified in the examples. We remark that the singularities of the Ruppeiner curvature are related to critical points in the thermodynamics (see for example, the recent work \cite{Mansoori} and references therein). We shall however, not be concerned with the curvature singularities in the work and rather focus on probing the interactions relying on the sign of the curvature.

\section{BTZ black holes}
Black hole solutions in \((2+1)\)-dimensional topological gravity were found by Banados, Teitelboim and Zanelli (BTZ) \cite{Banados:1992wn,Banados:1992gq}. These are the BTZ black holes with a negative cosmological constant which from the black hole thermodynamics perspective, leads to a positive thermodynamic pressure and therefore a fluidlike behavior in extended black hole thermodynamics. Even though in \((2+1)\)-dimensions gravitational fields do not have dynamical degrees of freedom and general relativity has no Newtonian limit, \((2+1)\)-dimensional theories are studied because they are often easier to work with yet share similar properties with their \((3+1)\)-dimensional counterparts. Despite certain similarities the BTZ black holes share with higher dimensional black holes in extended black hole thermodynamics, these \((2+1)\)-dimensional black holes are associated with some peculiar thermodynamic behavior. For instance, static\footnote{With static, we mean \(Q=0, J=0\).} BTZ black holes behave exactly like an ideal gas if the specific volume of the black hole is understood as the fluid volume. This feature is unlike any other asymptotically AdS black holes in higher dimensions which approach this ideal gas behavior only when the limit of large \(r_+\) is taken. Another important feature is that BTZ black holes do not admit any critical behavior unlike most higher dimensional counterparts which show critical behavior just like non-ideal systems in thermodynamics. BTZ black holes are therefore interesting to study in their own right apart from being regarded as toy models for asymptotically AdS black holes in higher dimensions.

\smallskip

The thermodynamics of static BTZ black holes is described by their enthalpy being given as~\cite{Dolan:2010ha},
\begin{equation}\label{BTZenthalpy}
 M = H(S,P) = \frac{4PS^2}{\pi},
\end{equation} where the black hole entropy is defined from the horizon radius \(r_+\) as,
\begin{equation}
  S = \frac{A}{4}, \hspace{3mm} A = 2\pi r_+.
\end{equation}
This leads to the following equation of state,
\begin{equation}
P\sqrt{V}=\frac{\sqrt{\pi}T}{4}.
\end{equation}
The equation of state exactly corresponds to the ideal gas limit of black holes in \(d=3\) where the specific volume in this case is identified as, \(v = 4\sqrt{V/\pi}\). All other black holes in \(d \geq 4\) approach this ideal gas behavior \(PV^{1/(d-1)} \sim T\) in the large \(r_+\) limit. A straightforward calculation shows that the Ruppeiner metric for this case is flat, i.e. the scalar curvature is zero as expected for the case of an ideal gas. We shall conclude from here that non-rotating and neutral BTZ black holes are associated with microstructures that do not interact.
%%%%%%%%%%%%%%%%%%%%%%%%%%%%%%%%%%%%%
\subsection{BTZ black holes with electric charge and angular momentum}\label{btznormal}
\begin{figure}[h]
%	 \begin{wrapfigure}{l}{0.3\textwidth}
	\begin{center}
		\centering
		\includegraphics[width=4.2in]{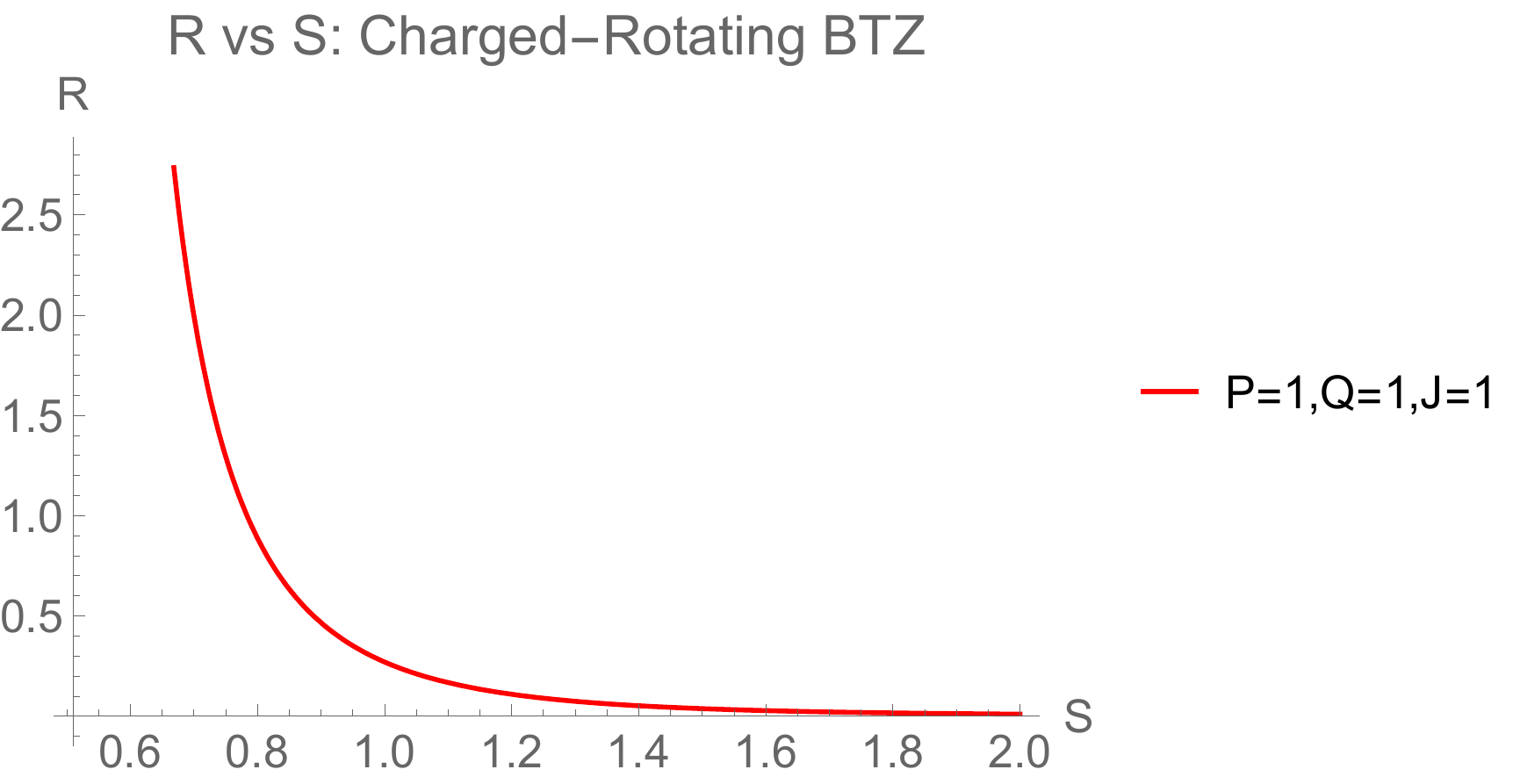}  		
		\caption{Ruppeiner curvature for a charged-rotating BTZ black hole on the \((S,P)\)-plane as a function of \(S\) for fixed \(P\).}
        \label{R_SP_S_JQ_excluding_T_negative_region}	
        \includegraphics[width=4.2in]{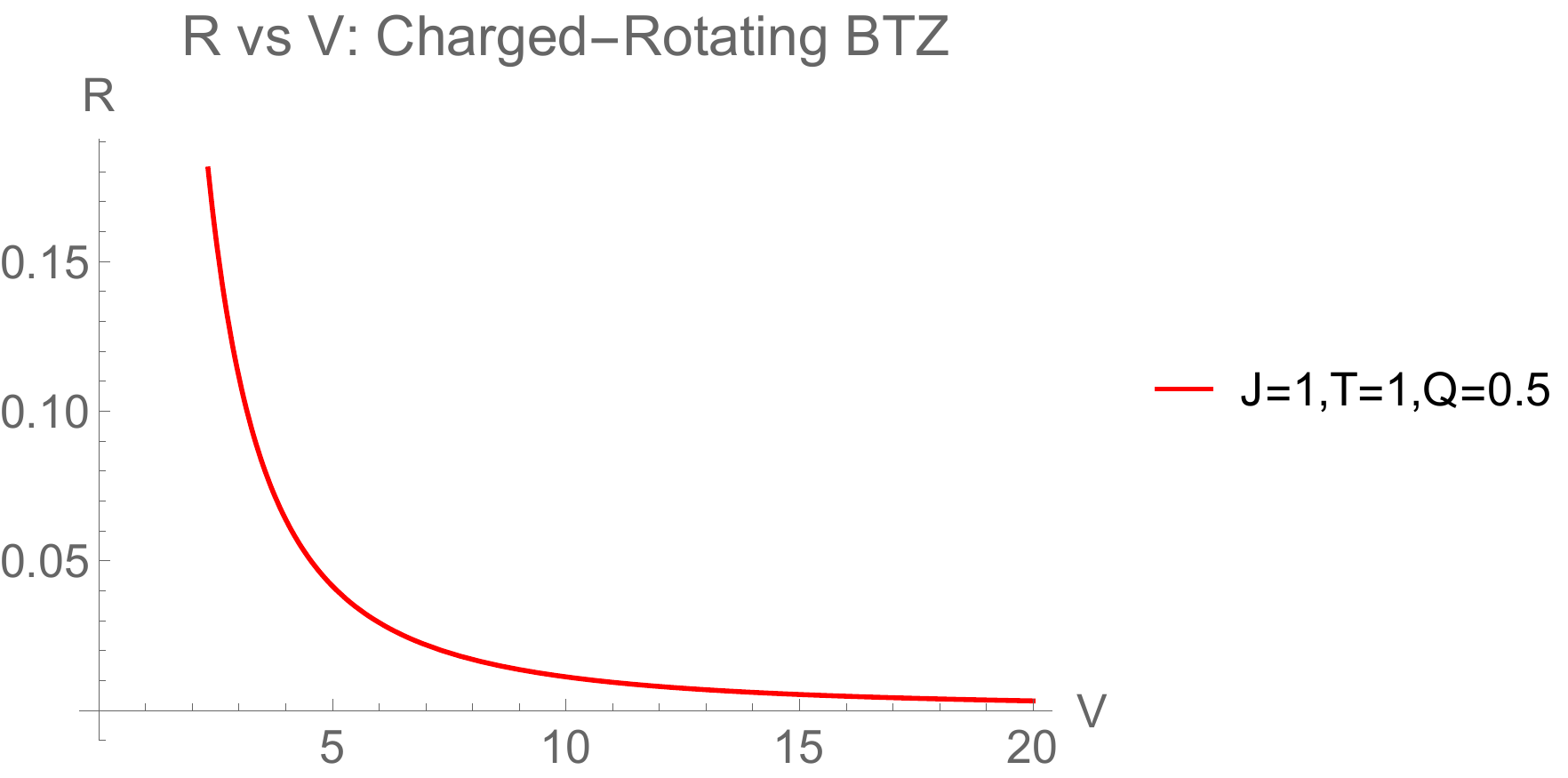}  		
		\caption{Ruppeiner curvature for a charged-rotating BTZ black hole on the \((T,V)\)-plane as a function of \(V\) for fixed \(T\).}
        \label{R_TV_V_JQ}	
	\end{center}
%	\end{wrapfigure}
\end{figure}

%%%%%%%%%%%%%%%%%%%%%%%%%%%%%%%%
We now turn to BTZ black holes which are associated with both electric charge and angular momentum. The thermodynamics is much more interesting and the corresponding behavior akin to that of a non-ideal fluid, i.e. one with interactions. The thermodynamics is however not of the van der Waals type as will be pointed out later. The enthalpy is given by \cite{Frassino:2015oca},
\begin{equation}\label{BTZQJenthalpy}
H(S,P) = \frac{\pi ^2 J^2}{128 S^2}-\frac{1}{32} Q^2 \log \left(\frac{32 P S^2}{\pi }\right)+\frac{4 P S^2}{\pi}.
\end{equation}
The thermodynamic volume now depends on charge and is not directly related to entropy. As a result, $C_V$ is not zero for charged BTZ black holes. The volume is given as,
\begin{equation}\label{BTZV}
V = \frac{4 S^2}{\pi }-\frac{Q^2}{32 P}.
\end{equation}
Specific heat \(C_P\) for the charged and rotating case is given by,
\begin{equation}
C_P = -\frac{\pi ^3 J^2 S-512 P S^5+4 \pi  Q^2 S^3}{3 \pi ^3 J^2+512 P S^4+4 \pi  Q^2 S^2}.
\end{equation}.
From eqn (\ref{BTZQJenthalpy}), it is possible to calculate the Ruppeiner metric on the \((S,P)\)-plane analytically, with the Ruppeiner curvature obtained to be,
\begin{eqnarray}\label{RBTZ}
R & =& \frac{-\pi  \left(A_1+A_2 +A_3\right)}{B}  \\
A_1 & =&3 \pi ^9 J^6 Q^2 \left(1280 P S^2-3 \pi  Q^2\right)+13824 \pi ^6 J^4 P Q^2 S^4 \left(\pi  Q^2-128 P S^2\right) \nonumber \\
   A_2 &=& 24576 P Q^2 S^8
   \left(\pi  Q^2-128 P S^2\right)^3 \nonumber \\
   A_3 &=&16 \pi ^2 J^2 S^4 \left(\pi  Q^2-128 P S^2\right)^2 \left(-262144 P^2 S^4+3072 \pi  P Q^2 S^2+\pi ^2 Q^4\right)\nonumber \\
   B &=& \left(\pi ^3 J^2 S-512 P S^5+4 \pi  Q^2 S^3\right) \left(3 \pi ^4 J^2 Q^2+4 S^2 \left(\pi  Q^2-128 P S^2\right) \left(256 P S^2+\pi
   Q^2\right)\right)^2. \nonumber
\end{eqnarray}
The curvature for the metric on the \((S,P)\)-plane is plotted in figure-(\ref{R_SP_S_JQ_excluding_T_negative_region}) as a function of \(S\) for fixed \(P\). As seen from figure-(\ref{Tr_BTZ}), temperature is positive only beyond $S=0.5$ and hence only that region in figure-(\ref{R_SP_S_JQ_excluding_T_negative_region}) is shown, signifying the presence of repulsive interactions.
We may alternatively calculate the Ruppeiner metric on the \((T,V)\)-plane using the Helmholtz free energy representation. The corresponding curvature is plotted as a function of \(V\) in figure-(\ref{R_TV_V_JQ}). It is clearly noted that the curvatures carry a positive sign in either of the planes clearly indicating repulsive interactions.
%However, the results are more transparent if we consider separate cases with either \(Q=0\) or \(J=0\) as will be done next.
%\begin{figure}[h]
%	 \begin{wrapfigure}{l}{0.3\textwidth}
%	\begin{center}
%		\centering
%		\includegraphics[width=4.6in]{R_TV_V_JQ}  		
%		\caption{Ruppeiner curvature for a charged-rotating BTZ black hole on the \((T,V)\)-plane as a function of \(V\) for fixed \(T\).}   \label{R_TV_V_JQ}		
%	\end{center}
%	\end{wrapfigure}
%\end{figure}
%\subsubsection{Charged non-rotating case}
\noindent
For the case with \(Q \neq 0\) and \(J = 0\), the specific heat at constant volume \(C_V\) is non-zero and is expressed as,
\begin{equation}
C_V = \frac{4 S^3-\pi  S V}{\pi  V-12 S^2},
\end{equation}
and the Ruppeiner curvature on the \((S,P)\)-plane can be obtained from eqn (\ref{RBTZ}) to be,
\begin{equation} \label{Qnonzero}
R_{J=0} = \frac{384 \pi  P Q^2 S}{\left(256 P S^2+\pi  Q^2\right)^2}.
\end{equation}
We note that charged black holes can be associated with electrically charged microstructures \cite{Wei:2015iwa,AR,Mann2019,Wei:2019yvs,GBRuppeiner} which are repulsive. This explains the origin of the positive scalar curvature of the Ruppeiner metric and hence, repulsive interactions between black hole microstructures from a phenomenological level.
It should also be remarked that charged BTZ black holes can be thought of as a particular case of a general fluid with the equation of state,
\begin{equation}\label{BTZfluideqnofstate}
P = \frac{T}{v} \, + \, \frac{a}{v^2} \, .
\end{equation} where \(a>0\) with the specific choice of the constant, \(a = Q^2/2\pi\) in eqn (\ref{BTZfluideqnofstate}) corresponding to the case of charged BTZ black hole. The equation of state resembles with that of the van der Waals fluid. The second term in the right hand side however, carries a positive sign as opposed to the negative sign for the case of the van der Waals fluid. This signifies the presence of repulsive interactions unlike the attractive interactions that are present in the van der Waals case. Moreover, the presence of such a repulsive term in eqn (\ref{BTZfluideqnofstate}) imposes a lower bound in the specific volume which for the case of the charged BTZ black hole is given by\footnote{In the present case, the thermodynamic volume $V$ and specific volume $v$ are not linearly related, as seen from eqn (\ref{BTZV}), i.e., $V =\frac{\pi  v^2}{16}-\frac{Q^2}{32 P} $. To get the correct $v_{\rm min}$, one has to choose $V_{\rm min}=0$. There is however an alternative possibility of retaining a linear relation between $V$ and $v$, at the cost of introducing a new renormalization length scale \cite{Frassino:2015oca}, in which case $v_{\rm min}=V_{\rm min}$. We however continue using the definition of $V$ as obtained in eqn (\ref{BTZV}) from enthalpy.}, \begin{equation}\label{vminQ}
v_{\rm min} = \frac{Q}{\sqrt{2 \pi } \sqrt{P}}.
\end{equation}
We now consider the neutral \((Q=0)\) and rotating case \((J \neq 0)\). We note that since the volume doesn't depend on \(J\), in this case the entropy and the volume are independent. Consequently, $C_V=0$ for neutral rotating BTZ black holes. From eqn (\ref{RBTZ}), we note the Ruppeiner curvature on the \((S,P)\)-plane for this case to be,
\begin{equation}
R_{Q=0} = \frac{4 \pi ^3 J^2}{512 P S^5 - \pi ^3 J^2 S}.
\end{equation}
The Ruppeiner curvature is always positive if the temperature \(T\) of the black hole is taken to be positive. On the \((T,V)\)-plane, the Ruppeiner curvature is even simpler, given as,
\begin{equation}
  R'_{Q=0} = \frac{J^2}{T V^2} \, .
\end{equation}
A few simple manipulations will show that the Ruppeiner curvatures on either of the planes are equivalent and are positive definite if we demand positivity of the temperature. The interactions are therefore repulsive. Since the neutral and non-rotating BTZ black holes are associated with microstructures that are non-interacting, it is then natural to think of the rotating cases as being associated with additional microstructures which interact repulsively and carry the degrees of freedom of the total angular momentum. On the other hand, if a black hole is associated with both electric charge and angular momentum, both classes of these repulsive microstructures are present thereby leading to an overall repulsion or equivalently a positive sign of Ruppeiner curvature, as obtained from eqn (\ref{RBTZ}).

\smallskip

The thermodynamic equation of state for the rotating BTZ black holes is given by,
\begin{equation}\label{RotatingBTZeqnofstate}
P = \frac{T}{v} \, + \, \frac{8J^2}{\pi v^4},
\end{equation}
where $v=4 r_+$ is the specific volume. The second term in the right hand side implies repulsive interactions and consequently no phase transitions associated with the rotating BTZ black hole. If the black hole were a fluid, such a term would account for repulsions between four molecules and appears in the virial expansion form of the equation of state for non-ideal fluids. In fact, eqn (\ref{RotatingBTZeqnofstate}) corresponds to a particular case of the RN-AdS fluid proposed recently in \cite{Bairagya:2019lxq} with the absence of the bimolecular attraction term. One notes the presence of a minimum volume being given by,
\begin{equation}\label{vminJ}
v_{min}=\frac{2^{3/4} \sqrt{J}}{\sqrt[4]{\pi } \sqrt[4]{P}}\, .
\end{equation}

%%%%%%%%%%%%%%%%%%%
\section{General exotic BTZ black holes}
%%%%%%%%%%%%%%%%%%%

General exotic BTZ black holes \cite{Townsend:2013ela,Frassino:2015oca,Frassino:2019fgr} which originate from gravitational actions following from purely Chern-Simons terms in three dimensions have generated a lot of interest, as being examples of systems which may be superentropic\footnote{They violate the Reverse Isoperimetric (RI) inequality~\cite{Cvetic:2010jb}, by having more entropy than that is allowed by RI inequality. For these systems a new instability conjectures are being actively pursued \cite{Johnson:2019mdp,Cong:2019bud,Johnson:2019wcq,Appels:2019vow}, which depend on the signs of $C_P$ as well as $C_V$}.  Let us start from the $(2+1)$-dimensional rotating BTZ black hole with metric given as,
\begin{equation} \label{exoticmetric}
    ds^2 = -V(r)dt^2+\frac{1}{V(r)}dr^2+r^2(d\theta-\frac{4 j}{r^2}dt)^2\, ,
\end{equation}
with
\begin{equation}
    V(r) = -8m+\frac{r^2}{l^2}+\frac{16j^2}{r^2}\,.
\end{equation}
The inner and outer horizon radii are given as: $r_{\pm} = 2\sqrt{l(l m\pm\sqrt{l^2m^2-j^2})}$. This form of the metric actually solves the Einstein equations with a negative cosmological constant in three dimensions in quite different cases \cite{Witten:1988hc}. More generally, if one takes the form of gravitational action to be of the form $I = \alpha \,  I_{EH}\, + \, \gamma I_{GCS}$, where $I_{EH}$ stands for the Einstein-Hilbert (EH) action and $ I_{GCS}$ gravitational Chern-Simons (GCS) action \cite{Deser:1981wh,Solodukhin:2005ah}, then it is possible to obtain more general situations where the form of the black hole metric does not change, but the parameters of the black hole satisfy novel relations \cite{Townsend:2013ela,Frassino:2015oca,Frassino:2019fgr}. For the case of an exotic black hole, the conserved mass $M$ and angular momentum $J$ are related to the parameters in the metric (\ref{exoticmetric}) as,
\begin{align}
\label{exoticMJ}
    M = \alpha m +\gamma \frac{j}{l}\, ,\\
    J = \alpha j+\gamma l m\, .
\end{align}
Here, $\alpha$ and $\gamma$ are constant coupling functions with limits: $\alpha\in[0,1]$ and
\begin{equation} \label{alphagamma}
\alpha + \gamma = 1 \,.
\end{equation} The case with $\alpha = 1$ corresponds to the standard rotating BTZ black hole~\cite{Carlip:1994hq}, where as $\alpha = 0$ leads to the purely exotic BTZ black hole~\cite{Townsend:2013ela}. More general situations are noted in other models~\cite{Carlip:1994hq,Carlip:1991zk,Banados:1997df,Banados:1998dc}. Mass $M$, angular momentum $J$ and entropy $S$ for general exotic BTZ black holes are given as \cite{Frassino:2019fgr},
\begin{align}
    M& = \frac{\alpha\big(r_-^2+r^2_+\big)}{8 l^2}+\frac{\gamma r_-r_+}{4 l^2} \label{M}\,, \\
        J& = \frac{\alpha r_-r_+}{4l}+\frac{\gamma\big(r_-^2+r^2_+\big)}{8l}\label{J}\,,\\
    S&= \frac{1}{2} (\pi\alpha r_+ +\pi\gamma r_-)\label{S}\,,\\
        V&= \alpha \pi r_+^2+\gamma \pi r_-^2\bigg(\frac{3r_+}{2r_-}-\frac{r_-}{2r_+}\bigg)\label{V}\,.
\end{align}
 The temperature and angular velocity are given as,
  \begin{align}
    T& = \frac{r_+^2-r^2_-}{2\pi l^2r_+} \,;\quad \Omega = \frac{r_-}{r_+ l}\label{T}\,,
    \end{align}
which are same for either normal BTZ ($\alpha=1$) or exotic BTZ ($\alpha=0$). Together, the above thermodynamic variables satisfy the first law, $dM = TdS+VdP+\Omega dJ$ and also respect the Smarr relation, $TS-2PV+\Omega J=0$. The specific choice $\alpha = 8/\pi^2$, reproduces the results of normal BTZ black holes discussed in previous sections. In this section, we concentrate on the cases where both $\alpha$ and $\gamma$ take different values restricted by the constraint , which corresponds to the general exotic BTZ black holes. The enthalpy can be obtained from mass $M$ in eqn (\ref{M}) by eliminating $r_-$ and $r_+$, in favor of thermodynamic variables to be,
\begin{align}\label{Hexoticbtz}
H(S,P) & =  \frac{2 \sqrt{P}}{\pi  (\alpha -1)^2} \left( 2\ 2^{3/4} (\alpha -1) S \sqrt{\frac{(1-2 \alpha ) \sqrt{P}}{\alpha -1}} \sqrt{-\frac{\pi ^{3/2} (\alpha -1) J+\sqrt{2} \sqrt{P} S^2}{\alpha -1}} \right. \nonumber \\
& \left.~~+ \sqrt{2} \pi ^{3/2} (\alpha -1) \alpha  J + 4 S^2 \sqrt{P} \alpha\right)\, ,
\end{align}
where for later convenience, $\gamma$ was eliminated using eqn (\ref{alphagamma}), with out loss of generality and thus we have $\alpha \leq 1$.
The  temperature $T$ and volume $V$, resulting from enthalpy in eqn (\ref{Hexoticbtz}) are given respectively as,
\begin{align}\label{Texotic}
& T = -\frac{4 \left(\frac{2^{3/4} \pi ^{3/2} (1-2 \alpha ) J P}{\sqrt{\frac{(1-2 \alpha ) \sqrt{P}}{\alpha -1}}}+4 P S \left(\sqrt[4]{2} S
   \sqrt{\frac{(1-2 \alpha ) \sqrt{P}}{\alpha -1}}-\alpha  \sqrt{-\pi ^{3/2} J-\frac{\sqrt{2} \sqrt{P} S^2}{\alpha -1}}\right)\right)}{\pi
   (\alpha -1)^2 \sqrt{-\pi ^{3/2} J-\frac{\sqrt{2} \sqrt{P} S^2}{\alpha -1}}},
\end{align}
\begin{align}\label{Vexotic}
& V = \frac{V_1+ V_2}{\pi  (\alpha -1)^3 \sqrt{P} \sqrt{\frac{(1-2 \alpha ) \sqrt{P}}{\alpha -1}} \sqrt{-\frac{\pi ^{3/2} (\alpha -1) J+\sqrt{2} \sqrt{P} S^2}{\alpha
   -1}}} \\
& V_1=  8 \sqrt{P} S^2 \left((\alpha -1) \alpha  \sqrt{\frac{(1-2 \alpha ) \sqrt{P}}{\alpha -1}} \sqrt{-\frac{\pi ^{3/2} (\alpha -1) J+\sqrt{2} \sqrt{P}
   S^2}{\alpha -1}}+\sqrt[4]{2} (2 \alpha -1) \sqrt{P} S\right) \nonumber \\
& V_2=\sqrt{2} \pi ^{3/2} (\alpha -1) J \left((\alpha -1) \alpha  \sqrt{\frac{(1-2 \alpha ) \sqrt{P}}{\alpha -1}} \sqrt{-\frac{\pi ^{3/2} (\alpha -1)
   J+\sqrt{2} \sqrt{P} S^2}{\alpha -1}}+3 \sqrt[4]{2} (2 \alpha -1) \sqrt{P} S\right).
\nonumber
\end{align}
Writing back in appropriate variables, it can of course be checked that these are same as the ones given in equations (\ref{V}) and (\ref{T}). Now, we note that it is in general difficult to write an equation of state $P=P(V,T)$ in the case of exotic BTZ black holes, due to the highly non-linear nature of equations  (\ref{Texotic}) and (\ref{Vexotic}), as indicated in~\cite{Frassino:2019fgr}.
\begin{figure}[h]
%	 \begin{wrapfigure}{l}{0.3\textwidth}
	\begin{center}
		\centering
		\includegraphics[width=5.2in]{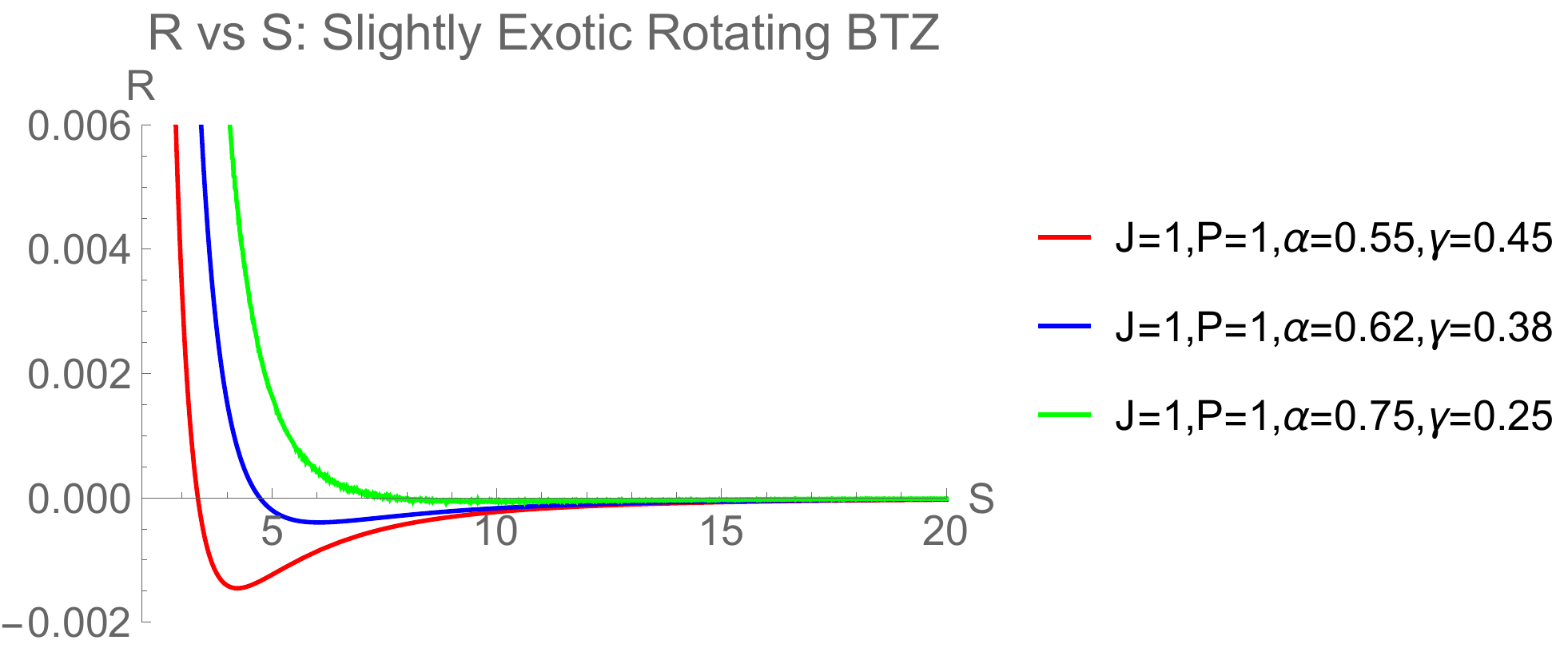}  		
		\caption{Ruppeiner curvature on \((S,P)\)-plane for slightly exotic rotating BTZ black holes as a function of \(S\) for fixed \(P\) and \(J\).}   \label{r_sp_s_j_slightly_exotic}		
%		\includegraphics[width=4.2in]{T_SP_S_J_slightly_exotic}  		
%		\caption{Temperature w.r.t. to Entropy for slightly exotic ($\alpha=3/4,\gamma=1/4$) rotating BTZ black holes.}   \label{T_SP_S_J_slightly_exotic}	
	\end{center}
%	\end{wrapfigure}
\end{figure}
%The Ruppeiner curvature also vanishes at the special point $S = \frac{\pi ^{3/4} \sqrt{\text{$\gamma $}} \sqrt{J}}{2^{3/4} \sqrt[4]{P}}$.
\begin{figure}[h]
%	 \begin{wrapfigure}{l}{0.3\textwidth}
	\begin{center}
		\centering
		\includegraphics[width=5.2in]{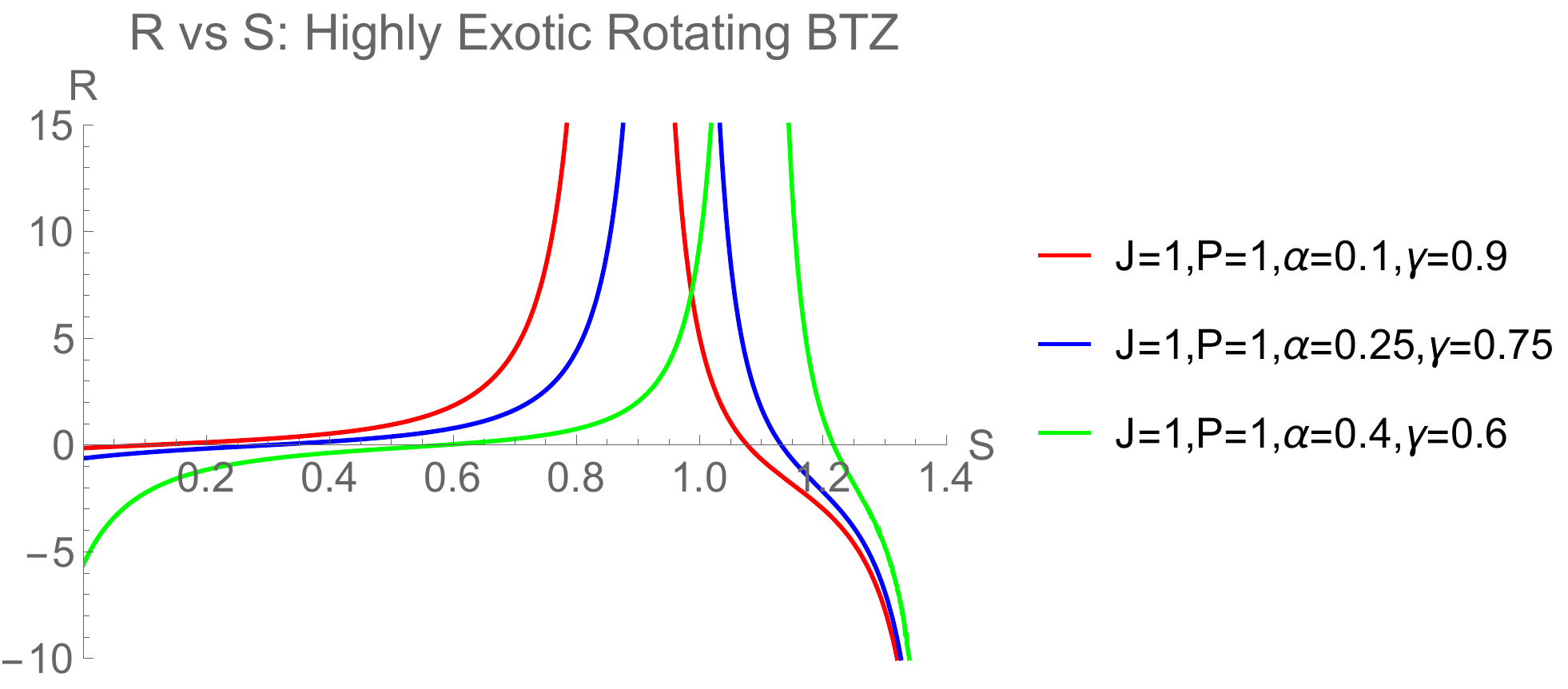}  		
		\caption{Ruppeiner curvature on \((S,P)\)-plane for highly exotic rotating BTZ black holes as a function of \(S\) for fixed \(P\) and \(J\).}   \label{r_sp_s_j_highly_exotic}	
%		\includegraphics[width=4.2in]{T_SP_S_J_highly_exotic}  		
%		\caption{Temperature w.r.t. to Entropy for slightly exotic ($\alpha=1/4,\gamma=3/4$) rotating BTZ black holes.}   \label{T_SP_S_J_highly_exotic}		
	\end{center}
%	\end{wrapfigure}
\end{figure}
Now, using the form of the metric in the $(S,P)$-plane in eqn (\ref{RuppeinerSP}), the Ruppeiner curvature can be calculated in the most general case analytically. The special case of purely exotic black hole ($\alpha=0, \gamma=1$) is treated separately and more details are given later in this section. For general exotic black holes, since the expressions are quite cumbersome we directly give the plots for Ruppeiner curvature, and classify the results in two broad classes, namely, $\alpha > \gamma$ (slightly exotic) and $\alpha < \gamma$ (highly exotic).  The thermodynamics and also phase structure of BTZ black holes in these two broad classes is different and hence need to be treated separately, as noted in~\cite{Frassino:2019fgr}. Figures-(\ref{r_sp_s_j_slightly_exotic}) and (\ref{r_sp_s_j_highly_exotic}), summarize the results of the computation of curvature for various values of $\alpha$ and $\gamma$. The most important feature, we find from these figures is that the Ruppeiner curvature crosses over from positive to negative at a valid point (meaning, the temperature is positive) of entropy and represents the presence of both attraction and repulsion dominated regions in exotic BTZ black holes. This generic feature is quite contrasting to the case $\alpha=1, \gamma=0$ of normal BTZ black holes (charged and/or rotating cases) discussed in section-(\ref{btznormal}), where the microstructures are only repulsive. \\

\noindent
We now discuss two special cases, corresponding to normal and purely exotic BTZ black holes, which are the extreme cases, coming from making the choices $\gamma=0$ and $\alpha=0$, respectively. The purely exotic case is quite interesting, as it was one of the first studied cases of exotic black holes where mass and angular momentum are essentially interchanged, as seen from eqns (\ref{exoticMJ}). In these extreme cases, the enthalpy and related thermodynamic quantities cannot be obtained from eqn (\ref{Hexoticbtz}), as the limits $\alpha=0$ and $\alpha=1$ $(\gamma=0)$ are singular. However, the enthalpy for each of these cases can be obtained directly from (\ref{M}) to be\footnote{The enthalpy in eqn (\ref{Hnormal}) corresponding to normal BTZ black hole can be seen to be same as the one obtained from eqn (\ref{BTZQJenthalpy}), for the choice $\alpha=1, Q=0$ and scaling $J$ by a factor of 64.},
\begin{align}
& H^{\rm normal}_{\gamma=0}\left(S,P \right) = \frac{\pi ^2 \alpha  J^2}{2 S^2}+\frac{4 P S^2}{\pi  \alpha }, \label{Hnormal}\\
& H^{\rm purely~exotic}_{\alpha=0}\left(S,P \right) = \frac{4 P S \sqrt{\frac{2 \sqrt{2} \pi ^{3/2} \gamma  J}{\sqrt{P}}-4 S^2}}{\pi  \gamma }. \label{Hpexotic}
\end{align}
For the case of normal rotating BTZ black holes, thermodynamic quantities can be derived from eqn (\ref{Hnormal}) and are noted to be same as in section-(\ref{btznormal}) (obtained by setting $Q=0$). For the case of normal BTZ black holes ($\gamma=0$), the Ruppeiner curvature is shown explicitly in eqn (\ref{Qnonzero}) and was found to be only positive. For the case of purely exotic rotating BTZ, the temperature and volume resulting from eqn (\ref{Hpexotic}) are,
\begin{align}
& T = \frac{8 \pi ^{3/2} \gamma  J \sqrt{P}-16 \sqrt{2} P S^2}{\pi  \gamma  \sqrt{\frac{\sqrt{2} \pi ^{3/2} \gamma  J}{\sqrt{P}}-2 S^2}}, \\
& V = \frac{6 \pi ^{3/2} \gamma  J S-8 \sqrt{2} \sqrt{P} S^3}{\pi  \gamma  \sqrt{P} \sqrt{\frac{\sqrt{2} \pi ^{3/2} \gamma  J}{\sqrt{P}}-2 S^2}}.
\end{align}
Now, using the above thermodynamic quantities in the form of the metric in the $(S,P)$-plane in eqn (\ref{RuppeinerSP}), the Ruppeiner curvature can be calculated\footnote{Ruppeiner curvature using a different three-dimensional metric was used in~\cite{Mohammadzadeh:2018zod} to study the critical points and phase transitions in exotic BTZ black holes} analytically to be,
%In fact, one can notice from figure-(\ref{R_SP_S_J_slightly_exotic}) for general exotic black holes, that as $\gamma \rightarrow 0$, the Ruppeiner curvature tends to be only positive, matching the observation in normal BTZ black holes. \\
\begin{align}
R_{{\rm\,purely\,exotic}}= \frac{4 \pi ^{3/2} \gamma  J \sqrt{P} S (A_1+A_2)}{B_1B_2} \, ,
\end{align}
where
\begin{align}
A_1& =-50320 \pi ^6 \gamma ^4 J^4 P^2 S^8+18336 \sqrt{2} \pi ^{9/2} \gamma ^3 J^3 P^{5/2} S^{10}+15744 \pi ^3 \gamma ^2 J^2 P^3
   S^{12} \nonumber \\
&~~~-18304 \sqrt{2} \pi ^{3/2} \gamma  J P^{7/2} S^{14}+9216 P^4 S^{16}\, ,\\
A_2&=-189 \pi ^{12} \gamma ^8 J^8+1827 \sqrt{2} \pi ^{21/2} \gamma ^7 J^7 \sqrt{P} S^2-13734 \pi ^9 \gamma ^6 J^6 P S^4+25938
   \sqrt{2} \pi ^{15/2} \gamma ^5 J^5 P^{3/2} S^6 \, , \\
B_1&=\sqrt{2} \pi ^3 \gamma ^2 J^2-6 \pi ^{3/2} \gamma  J \sqrt{P} S^2+4 \sqrt{2} P S^4 \, ,\\
 B_2&=\left(9 \pi ^6 \gamma ^4 J^4-81 \sqrt{2} \pi ^{9/2} \gamma ^3 J^3 \sqrt{P} S^2+408 \pi ^3 \gamma ^2 J^2 P S^4-392 \sqrt{2} \pi
   ^{3/2} \gamma  J P^{3/2} S^6+256 P^2 S^8\right)^2\, .
\end{align}
$R_{{\rm\,purely\,exotic}}$ is plotted below in figure-(\ref{r_sp_s_j_normal_vs_pexotic}) together with the Ruppeiner curvature for normal rotating BTZ black holes. Note that we can use the positivity of temperature from the relevant curves in figure-(\ref{r_sp_s_j_normal_vs_pexotic}) to find the physically meaningful range of entropy; as the temperature becomes negative for higher values of $S$ (purely exotic case) and there are also further thermodynamic instabilities~\cite{Cong:2019bud}.
\begin{figure}[h]
%	 \begin{wrapfigure}{l}{0.3\textwidth}
	\begin{center}
		\centering
		\includegraphics[width=5.5in]{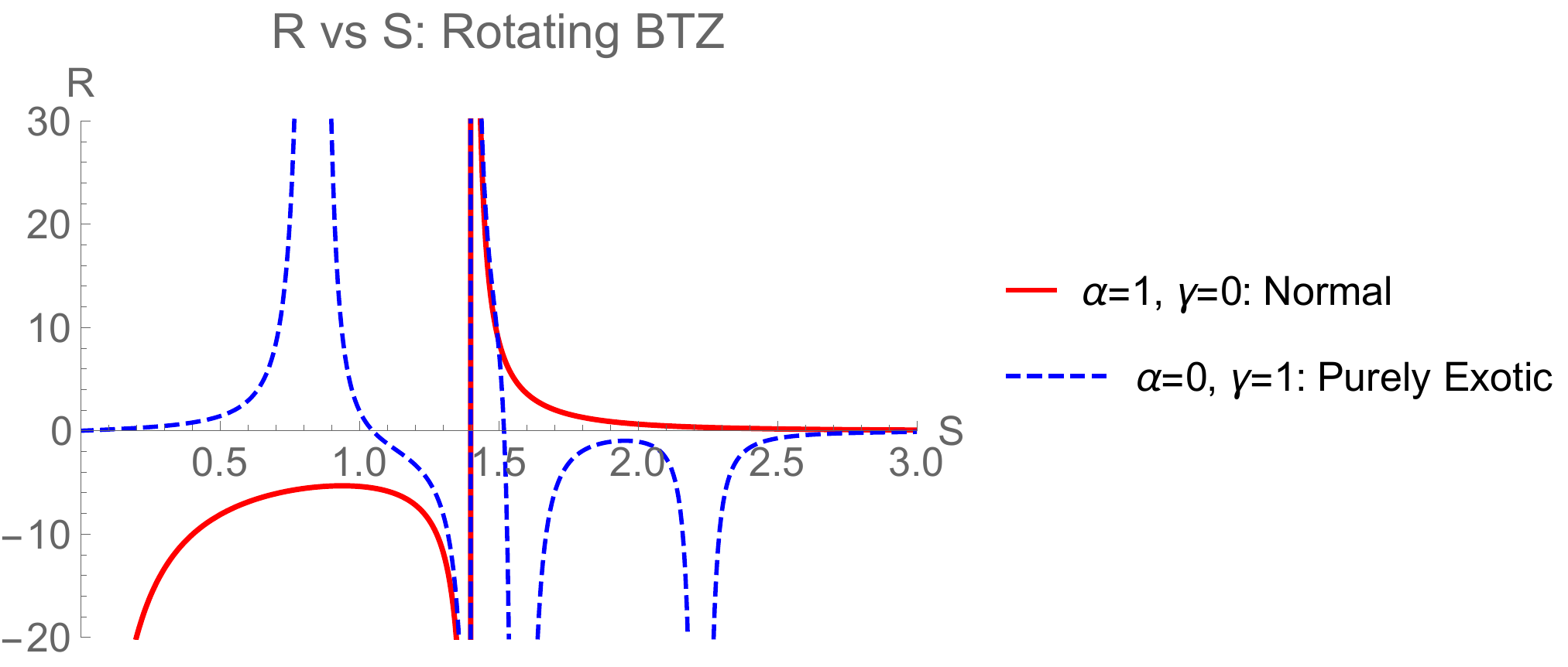}  		
		\caption{Ruppeiner curvature on \((S,P)\)-plane  for \(P=1\) and \(J=1\) for (a) Thick red curve: Normal rotating BTZ black holes ($\alpha=1,\gamma=0$) (b) Blue dashed curve: purely exotic rotating BTZ black holes ($\alpha=0,\gamma=1$).}   \label{r_sp_s_j_normal_vs_pexotic}
%		\includegraphics[width=4.2in]{T_SP_S_J_purely_exotic}  		
%		\caption{Temperature w.r.t. to Entropy for purely exotic rotating BTZ black holes.}   \label{T_SP_S_J_purely_exotic}		
	\end{center}
%	\end{wrapfigure}
\end{figure}
We note the following from figure-(\ref{r_sp_s_j_normal_vs_pexotic}):
For the normal BTZ black holes, the curvature is always positive and the plots are valid only beyond $S=1.403$ (The point before which temperature is negative, as seen from case of normal BTZ in figure-(\ref{T_SP_S_J_exotic})). On the other hand, for the purely exotic case,  the plots beyond the point $S=1.403$ are not valid, as the temperature becomes negative after this (as seen from purely exotic case in figure-(\ref{T_SP_S_J_exotic})). Furthermore, the crossing of Ruppeiner curvature from positive to negative at $S=1.04$ (for the particular choice of parameters shown in the figure), represents the shifting of dominant interactions from being repulsive to attractive.  It is of course important to remember that in the case of purely exotic BTZ black hole where $\alpha=0$, $\gamma=1$, the entropy is given in terms of the inner, rather than the outer horizon \cite{Townsend:2013ela}, providing a novel view point to the question of location of microstructures in black holes. \\

\noindent
The sign change of the Ruppeiner curvature indicates that both attraction and repulsion dominated regions exist for all the cases of general exotic BTZ black holes. The physics of such shifting of interactions at the point where the curvature scalar has a zero crossing can be understood qualitatively from the two fluid model \cite{AR,GBRuppeiner} which essentially considers such an interacting system as a binary mixture of two fluids respectively with purely repulsive and attractive interactions among the molecules which share the degrees of freedom of the total entropy of the system. The crossing point is then determined from the relative number densities of the fluid molecules of the two types which compete with one another deciding as to which kind of interaction has an overall dominance at a particular thermodynamic point. The number density has earlier been used to understand microstructures and phase transitions in charged black holes in AdS~\cite{Wei:2015iwa}. It should also be remarked that a change of sign of the Ruppeiner curvature does not imply a phase transition and is consistent with the previously noted result that exotic BTZ black holes do not admit a critical behavior \cite{Frassino:2019fgr}. The nature of microstructure interactions can some times be modeled, in the mean field approximation using an interaction potential between two neighboring fluid molecules. For instance, for the van der Waals fluid, interaction potential is given by the Lennard-Jones potential, which describes a short-range repulsive and a long-range attractive interaction. The attractive part is known to dominate in this system. But, repulsive interactions could still exist due to thermal fluctuation related effects or also inter molecular collisions (which is very much the case for charged black holes in AdS \cite{Mann2019}). It would be interesting to explore the aforementioned issues to gain a better understanding of microstructures of black holes.\\

%%%%%%%%%%%%%%%%%%%%%%%%%%%%%%
\section{Remarks}
\begin{figure}[h]
%	 \begin{wrapfigure}{l}{0.3\textwidth}
	\begin{center}
		\centering
		\includegraphics[width=5.2in]{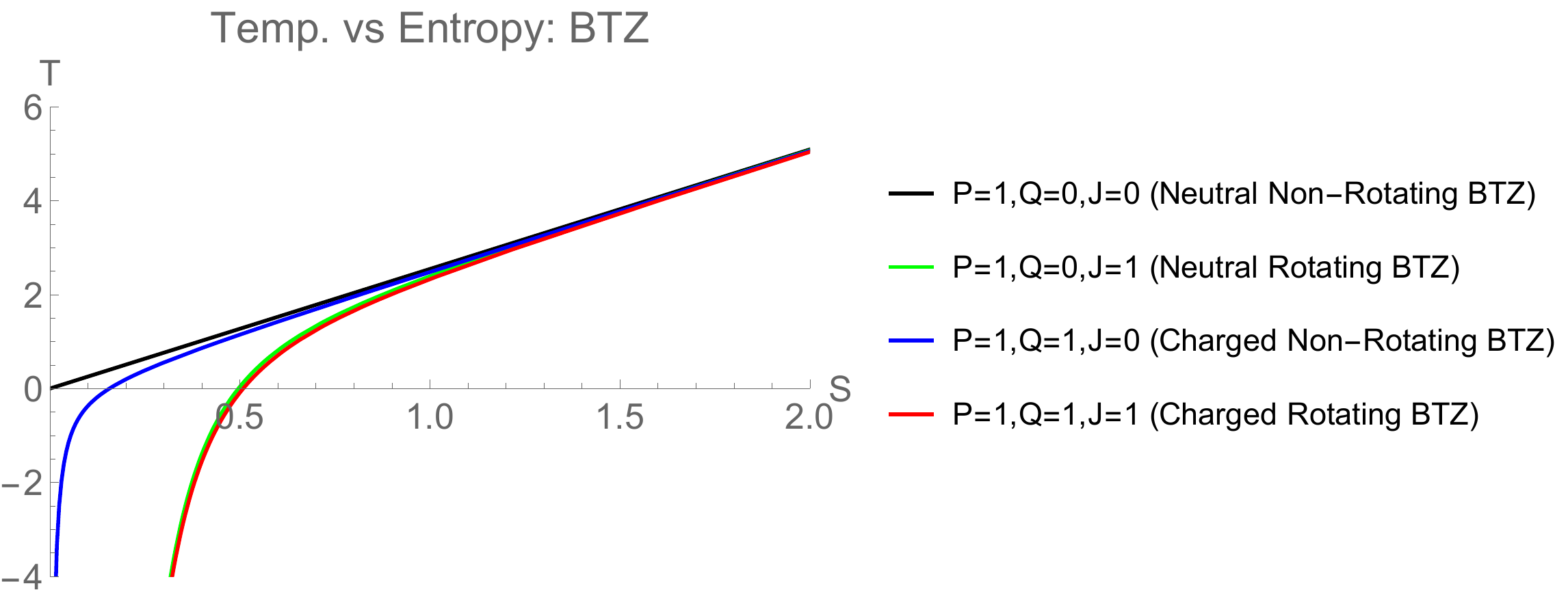}  		
		\caption{Variation of the black hole temperature \(T\) vs entropy \(S\) for BTZ black holes.}   \label{Tr_BTZ}
    \includegraphics[width=5.2in]{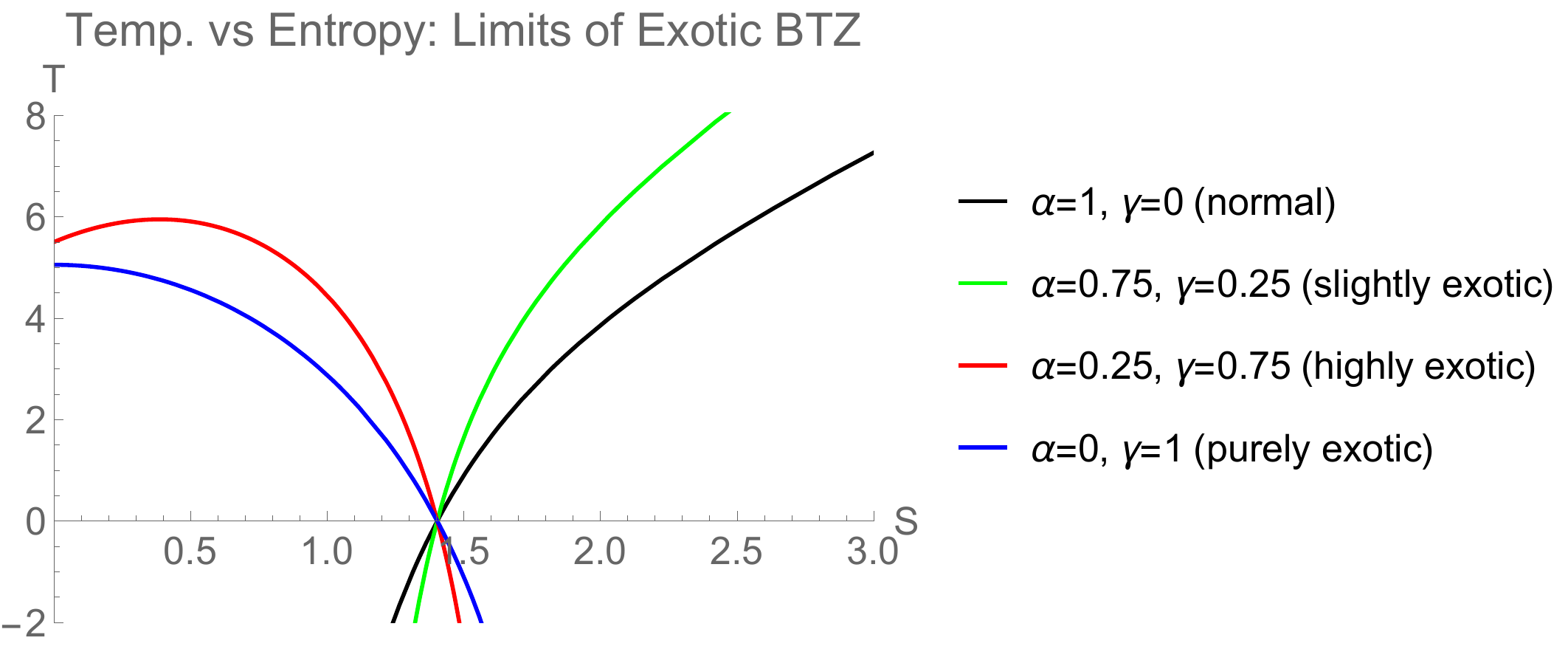}  		
		\caption{Temperature \(T\) with respect to entropy in several limits of exotic BTZ black holes. The behavior of curves for slightly exotic and highly exotic cases is similar for any other values of $\alpha$ and $\gamma$.}   \label{T_SP_S_J_exotic}	
	\end{center}
%	\end{wrapfigure}
\end{figure}

%%%%%%%%%%%%%%%%%%%%%%%%%%%%%%%%%
In this work, we have used Ruppeiner geometry to probe the nature of interactions between microstructures in the case of normal and exotic BTZ black holes. For the normal BTZ black hole, where \((\alpha=1,\gamma=0)\), it was recorded that the interactions are always repulsive for black holes carrying charge and/or angular momentum whereas there are no interactions between the microstructures for BTZ black holes without electric charge and/or rotation. In the case without charge or angular momentum, the BTZ black holes admit the thermodynamics of the ideal gas and as a result, the underlying microstructures are non-interacting. Moreover, at the thermodynamic level the similarity between angular momentum and electric charge of a black hole is well known in black hole thermodynamics. For instance, plotting the expression for temperature of non-exotic BTZ black holes,
\begin{equation}
T = -\frac{\pi ^2 J^2}{64 S^3}+\frac{8 P S}{\pi }-\frac{Q^2}{16 S},
\end{equation} where \(S = \pi r_+/2\) is the entropy of the black hole horizon shows that qualitatively, rotating and charged BTZ black holes behave in a similar fashion. The plot is shown in figure-(\ref{Tr_BTZ}). Therefore, similar to electrically charged microstructures for charged black holes as discussed in~\cite{Wei:2015iwa,AR,Mann2019,Wei:2019yvs,GBRuppeiner}, it is suggestive to associate rotating black holes with microstructures of another type, which carry the rotating degrees of freedom and interact with each other in a repulsive manner. For black holes with both electric charge and rotation, the interactions can add up since they are both repulsive. This also means that non-exotic BTZ black holes do not admit any critical behavior as has been noted earlier~\cite{Gunasekaran:2012dq}.
It is also interesting to make contact with the fluid picture proposed in~\cite{Bairagya:2019lxq}, with a novel equation state in a virial expansion (analogous, yet different from van der Waals fluids), which resembles the charged fluids in AdS and in particular, has non-zero heat capacity at constant volume~\cite{Landau}. This BTZ fluid picture is important as it can lead to a statistical mechanical understanding of the microstructures~\cite{RAJP}.  In fact, it is possible to have such a fluid picture, by proposing an equation of state as,
\begin{equation}\label{BTZfluid}
P = \frac{k_B\, T}{v_f} + \frac{a}{v_f^2} + \frac{d}{v_f^4} \, ,
\end{equation}
which resembles the charged and rotating BTZ black hole, only for the specific choice $a=Q^2$ and $d=J^2$. Here $v_f$ is the volume per molecule, and if $a$ and $d$ are arbitrary positive constants, eqn (\ref{BTZfluid}) stands for a general BTZ fluid with $C_{v_f} \neq 0$ in general. There is in fact a minimum volume for the fluid molecules, which can be found out by imposing the positivity of temperature in eqn (\ref{BTZfluid}) to be,
\begin{equation} \label{vmin}
v_{f,min} = \frac{1}{\sqrt{2P}}\left[\left(a^2+4Pd\right)^{1/2} + a\right]^{1/2} \, ,
\end{equation}
which differs from that of the RN-AdS fluid~\cite{Bairagya:2019lxq} in the sign of $a$. The minimum volumes noted in the special cases for only charged and only rotating BTZ black holes given in eqns (\ref{vminJ}) and (\ref{vminQ}), respectively, match with the expression given in eqn (\ref{vmin}), in appropriate limits. There is of course an important difference when compared to RN AdS fluids~\cite{Bairagya:2019lxq}. In the limit $P \to \infty$, minimum volume in eqn (\ref{vmin}) goes to zero, similar to the RN AdS fluids. But, at $P=0$, minimum volume in eqn (\ref{vmin}) goes to infinity. The thermodynamics is therefore, as one could appropriately call it, the anti-van der Waals type. An immediate consequence of this fact is that there are no phase transitions because such an anti-van der Waals fluid can never undergo a liquefaction due to presence of only repulsive interactions between molecules. It would be interesting to compute the Ruppeiner curvature for the BTZ fluid taking $C_{v_f}$ to be a constant~\cite{Mann2019,Wei:2019yvs,GBRuppeiner,Bairagya:2019lxq} and compare it with the one for BTZ black hole. The Ruppeiner curvatures in both cases may in general be different, but features, such as, presence of only repulsive interactions, are not expected to change for BTZ fluids.\\

\noindent
For the general exotic BTZ black holes, the Ruppeiner curvature was calculated exactly for various values of the parameters $\alpha$ and $\gamma$ and the results were presented by classifying them in to two broad classes, namely, $\alpha > \gamma$ (slightly exotic) and $\alpha < \gamma$ (highly exotic), apart from the case $\alpha=0, \gamma=1$, corresponding to purely exotic black holes. For, all three cases, it was noted that there are in general attraction and repulsion dominated regions, as the Ruppeiner curvature crosses zero, as noted from figures-(\ref{r_sp_s_j_slightly_exotic}), (\ref{r_sp_s_j_highly_exotic}) and (\ref{r_sp_s_j_normal_vs_pexotic}). We also found that the behavior of Ruppeiner curvature in figures-(\ref{r_sp_s_j_highly_exotic}) and (\ref{r_sp_s_j_normal_vs_pexotic}), for the highly exotic and purely exotic cases is similar, in which cases, the nature of temperature curves is also identical, as shown in figure-(\ref{T_SP_S_J_exotic})\footnote{$J$ in the curves used in figure-(\ref{T_SP_S_J_exotic}) is scaled by a factor of $64$, as compared to the one in figure-(\ref{Tr_BTZ})}. For the slightly exotic BTZ case, although, the temperature variation with respect to entropy is similar to that of normal BTZ (as noted from figure-(\ref{T_SP_S_J_exotic})), the behavior of Ruppeiner curvature is different. A change in sign of curvature happens for slightly exotic BTZ black holes as well, as seen from figure-(\ref{r_sp_s_j_slightly_exotic}), which should be compared with the appropriate curve in figure-(\ref{r_sp_s_j_normal_vs_pexotic}) for the standard rotating BTZ black hole, where such a sign change does not happen.  It would be interesting to explore other black hole systems in three dimensions, with the inclusion of dilaton and other non-linear couplings involving non-linear electromagnetic fields, where further interesting behavior of microstructures may exist, including novel critical behavior. More importantly, a statistical mechanical understanding of the microscopic degrees of freedom may be explored with phenomenological models for normal~\cite{Cai:1998ep} as well as exotic BTZ black holes~\cite{Frassino:2015oca}.\\

\end{document}